\documentclass[journal]{IEEEtran}
\usepackage{graphicx}
\usepackage{amsmath}
\usepackage[mathscr]{eucal}
\usepackage{amssymb}
\newcommand{\be}[1]{\begin{equation} \label{#1} }
\newcommand{\bea}[1]{\begin{eqnarray} \label{#1} }

\newcommand{\bfi}{\begin{figure}}
\newcommand{\efi}{\end{figure}} 
\newcommand{\ee}{\end{equation}}
\newcommand{\eea}{\end{eqnarray}}

\newcommand{\w}{{\omega}}

\begin{document}
\title{Electrically Small Supergain Endfire Arrays}
\author{Arthur D. Yaghjian,~\IEEEmembership{Fellow,~IEEE}, Terry H. O'Donnell,~\IEEEmembership{Member,~IEEE}, Edward E. Altshuler,~\IEEEmembership{Life Fellow,~IEEE}, and Steven R. Best,~\IEEEmembership{Fellow,~IEEE}
\thanks{The authors are with the Air Force Research Laboratory, Sensors Directorate, Hanscom AFB, MA 01731-2909 USA.  This work was supported by the U.S. Air Force Office of Scientific Research (AFOSR).}}
\maketitle
\begin{abstract} 
The theory, computer simulations, and experimental measurements are presented for {\em electrically small} two-element supergain arrays with near optimal endfire gains of  7 dB.  We show how the difficulties of narrow tolerances, large mismatches, low radiation efficiencies, and reduced scattering of electrically small parasitic elements are overcome by using electrically small {\em resonant} antennas as the elements in both separately driven and singly driven (parasitic) two-element electrically small supergain endfire arrays. Although rapidly increasing narrow tolerances prevent the practical realization of the maximum theoretically possible endfire gain of electrically small arrays with many elements, the theory and preliminary numerical simulations indicate that near maximum supergains are also achievable in practice for electrically small arrays with three (and possibly more) resonant elements if the decreasing bandwidth with increasing number of elements can be tolerated. 
\end{abstract}
\par
{\small\bf\em Index Terms}---{\small\bf  Supergain endfire arrays, resonant antennas, electrically small antennas.}\\
\section{Introduction}\label{sec I}
\PARstart{I}{n} his 1947 paper on the fundamental limitations of small antennas, Wheeler \cite{Wheeler-1947} defined a small antenna as ``one whose maximum dimension is less than the `radian-length' [$\lambda/(2\pi)$]," where $\lambda$ is the free-space wavelength. If one takes the radius $a$ of the sphere that circumscribes an antenna as its ``maximum dimension" measured from its center, then an antenna is electrically small if $ka <1$, where $k=2\pi/\lambda$ denotes the free-space wavenumber.\footnote{In later papers, for example, \cite{Wheeler-1959}, Wheeler defined a small antenna as one with $ka\ll 1$.  Also, Best \cite{Best-June2003} suggests the definition of a small antenna as $ka < 0.5$ based on how small a number of different open-ended, bent-wire antennas have to become for their radiation resistances to be approximately equal.  Here in the present paper, however, we use the less stringent criterion $ka<1$ as the definition of  an electrically small antenna because we shall be applying this criterion to array antennas with two or more elements.}  Since Wheeler's 1947 paper, a myriad of different electrically small antennas have been designed for a variety of applications \cite{Fujimoto} and, as a perusal of the issues of antenna journals indicates, the appearance of new designs and applications has continually accelerated \cite{Hansen-book}, \cite{Miron}.
\par
Yet, to our knowledge, none of these electrically small antennas have measured gains appreciably greater than the $10\log_{10}(1.5) = 1.76$ dB directivity of an elementary electric or magnetic dipole \cite[tbl. 6-2]{Kraus-book} --- even though early papers \cite{Oseen}--\cite{Riblet} (see \cite{Bloch-1960} for others) discussed the theoretical possibility of unlimited superdirectivity from arbitrarily small source regions, and it was shown by Uzkov in a 1946 paper \cite{Uzkov} that the endfire directivity of $N$ collinear isotropic radiators, each excited with the proper magnitude and phase, approaches a value of $N^2$ as the separation distance of the radiating elements approaches zero.  For example, a two-, three-, or four-element electrically small endfire array of isotropic radiators can, in principle, attain an endfire gain of 6.0 dB, 9.5 dB, or 12.0 dB, respectively.  Moreover,  for elementary dipole radiating elements, these theoretically possible maximum gains of two-, three-, and four-element electrically small endfire arrays increase to 7.2 dB, 10.3 dB, and 12.6 dB, respectively, and to even slightly higher gains for half-wavelength dipoles; see \cite[app. A and fig. 5]{Altshuler-2005} and Fig. \ref{fig4} below. It should be noted that \cite{Altshuler-2005} considered the excitation and measurement of quarter-wavelength monopoles (half-wavelength dipoles in free space) but did not address the design, construction, or measurement of electrically small antennas.
\par
A gain of $N^2$ represents a remarkable ``supergain" compared to the maximum possible gain, $N$, for isotropic radiators spaced a half wavelength apart \cite[eq. (A.16)]{Altshuler-2005}, especially because this supergain is attained as the length of the collinear array approaches zero.  And, in fact, it is not feasible to obtain close to this $N^2$ maximum endfire directivity in practice for a large number of elements because the required accuracy in the values of the magnitude and phase of the excitation currents increases very rapidly with the number of array elements $N$ \cite{Yaru}.  This is true even for array elements like half-wavelength dipoles that are not electrically small in the dimension normal to the endfire direction or axis of the array antenna.  However, for closely spaced ($\le 0.2 \lambda$ element spacing) two-, three-, and four-element endfire arrays of  nominally half-wavelength dipoles, free-space gains of approximately 6.7 dB, 9.2 dB, and 10.8 dB have been measured, respectively,  with two separately driven elements \cite{Altshuler-2005}, \cite{Best-July2005}, \cite{Lancaster-1992}, with three-element Yagi antennas (center element driven) \cite{NBSTN688}, and with four separately driven elements \cite{Bloch-1953}.  Closely spaced, two-element, half-wavelength dipole Yagi antennas with measured gains as high as 6 to 7 dB are commercially available \cite{Steppir} and two half-wavelength dipoles with equal but opposite currents and spaced about $\lambda/8$ or less (the W8JK array) achieve a gain of about 6 dB \cite{Brown}, \cite[pp. 185--186]{Kraus-book}. Closely spaced, three-element, meander-line ``Yagi-Uda arrays" with about 7.5 dB gain have been designed recently (though not constructed) with element heights of about a quarter wavelength \cite{Werner}; and closely spaced, three-element, half-wavelength ``folded Yagi arrays" with about 7 dB gain have been recently designed and measured \cite{Ling-AWPL}.  Also, closely spaced, single-feed, three-element patch antennas  approximately one wavelength across have been designed that have a few dB of gain at GHz frequencies \cite{Montrose-Popovic}.
\par
In contrast to these examples of supergain array antennas consisting of two, three, and four closely spaced $\lambda/2$ elements, electrically small ($ka<1$) two or three (or more) element array antennas with supergains reasonably close (within a dB or so) to the theoretical maximum have eluded practical realization. (It was shown in \cite{Nyquist} that an electrically small two-element parasitic array of impedance-loaded short wires exhibits a theoretical directivity of a few dB, and in \cite{Newman} that significant superdirectivity can be obtained in principle from an electrically small array of four separately driven short wires.  However, large mismatches and low radiation resistances prevented the practical realization of significant supergain from these arrays.)  Thus, the 1995 statement of Haviland \cite{Haviland} that ``the small high-gain and highly directive beam antenna remains `pie in the sky',"  describes the present state of the art of electrically small supergain antennas (including endfire antenna arrays with separately fed elements and parasitic arrays with a single feed element combined with parasitic elements).  It also sets the stage for the present paper in which we document the theory, computer simulations, and experimental measurements for separately fed and singly fed (parasitic) two-element electrically small supergain endfire arrays with free-space gains of 6 to 7 dB.  We also include the theoretically determined tolerance curves %and computer simulations
for three-element separately fed 
%and singly fed (parasitic) 
supergain endfire arrays.  The salient features of two-element, electrically small parasitic arrays were first presented in \cite{O&Y}.
\subsection{Electrically Small Resonant Elements}
\par
As we shall explain in more detail in Section \ref{sec II}, the key idea that enables the practical realization of electrically small supergain endfire arrays is the use of resonant antennas for the array elements.  By definition, resonant antennas (whether or not they are electrically small) have zero input reactance at their resonant frequencies.  Thus, as two identical electrically small resonant antennas are brought closer together within a small fraction of a wavelength to produce supergain, their input reactances are much smaller (in magnitude) than the input reactances of below-resonance electrically small electric-dipole antennas (which have high capacitive reactances).  These lower input reactances allow the array elements to be fed without the use of large tuning reactances that can add to the size and loss of the array antenna \cite{Newman}. (Although below-resonance electrically small wire-loop magnetic-dipole antenna elements have low (inductive) reactances, their extremely low radiation resistances preclude their use as elements in efficient arrays.) 
\par
A second advantage of using resonant antennas for the electrically small elements in a supergain array has to do with the lowering  of the radiation resistance of two elements as they are brought closer together to produce supergain \cite{Best-July2005}.  A lower radiation resistance implies a lower radiation efficiency, which reduces the gain proportionately, and requires a more sophisticated matching network to feed the array.  Electrically small resonant elements, however, can be designed with multiple arms  \cite{Dobbins&Rogers}--\cite{Lim-Ling-2006} that increase both the radiation resistance and efficiency,  and/or with parallel loops that increase the input resistance (though not the efficiency) \cite{Best-January2005}--\cite{Lim-Ling-2004}.
\par
Probably the most striking advantage of using resonant elements comes from the discovery that a resonant electrically small element with its input terminals shorted  behaves as an effective passive director or reflector (unlike a below-resonance electrically small shorted element) --- to the degree that an electrically small two-element supergain array with one element fed and one shorted parasitic element exhibits a supergain within a few tenths of a dB of the maximum possible supergain of the corresponding doubly fed two-element array \cite{O&Y}.  Moreover, this result appears to hold generally for all resonant antenna elements, and thus opens the possibility of a variety of single-feed, electrically small, parasitic supergain arrays.
\section{Fundamentals of Electrically Small Supergain Arrays}\label{sec II}
Five commonly stated or assumed reasons for the lack of progress in the development of electrically small supergain arrays can be summarized as follows.
\begin{enumerate}
\item The required tolerances on the magnitude and phase of the element input excitations are too tight to be maintained in practice.
\item Closely spaced electrically small elements have such high input reactances and such low radiation resistances that mismatch losses between the power supply and the antenna elements would prevent the practical realization of supergain. 
\item Even if the mismatch losses can be overcome with a well designed matching network, the ohmic losses in the electrically small elements and the matching network would dominate the low radiation resistance of the array antenna and eliminate any substantial supergain.  In other words, the radiation efficiency of an electrically small array would be too low to allow for supergain. 
\item Parasitic endfire arrays, the Yagi being the prime example, are attractive because they have just one fed element.  Yet, they are unsuitable for electrically small supergain endfire arrays since electrically small parasitic elements, unlike half-wavelength parasitic elements, would not make effective enough scatterers (reflectors or directors) to produce supergain.
\item The bandwidth of many electrically small supergain arrays would be too narrow for many applications.
\end{enumerate}
In regard to the bandwidth concerns expressed in 5), all electrically small antennas have quality factors ($Q$s) that are larger (and usually many times larger) than $0.5/(ka)^3$ \cite{Yaghjian(Felsen-paper)}, and thus are narrow-band for $ka \ll 1$ unless they are fed through complex tuning circuits or are specially designed to have multiple resonances at closely spaced frequencies.  Unfortunately, widening the bandwidth with complex tuning circuits and special designs for multiple resonances is generally not compatible with low loss and keeping the entire antenna system electrically small at GHz frequencies.  Moreover, as two electrically small antenna elements are brought closer together than a half-wavelength, the radiation resistance decreases, the $Q$ increases and the bandwidth decreases (typically by a factor of about five at $\lambda /8$ spacing).
\par
In this paper we shall avoid the legitimate bandwidth concerns of 5) by simply restricting ourselves to narrow band applications.  Still, we have to overcome the difficulties expressed in reasons 1) -- 4), namely, tight tolerances, large mismatches, low radiation efficiency, and reduced scattering of electrically small parasitic elements.  
\subsection{Tolerances}
The required tolerances on the magnitudes and phases of the element excitations for closely spaced electrically small supergain endfire arrays should be comparable to those for supergain endfire arrays with similarly spaced half-wavelength electric dipole elements.  Since, as mentioned in the Introduction, two-, three-, and four-element supergain endfire arrays have been constructed with closely spaced half-wavelength dipoles,
% \cite{Altshuler-2005}, \cite{Best-July2005},  \cite{Lancaster-1992}, \cite[fig. 5(A)]{ARRL}, \cite{Ling},  \cite{Bloch-1953}
the tolerances for electrically small supergain arrays with just a few elements should not be prohibitive.  Moreover, Newman and Schrote \cite{Newman} have shown that the tolerance constraints on an electrically small four-element array of separately driven short wires are not too restrictive to prevent the realization of supergain.  (They did not achieve supergain in practice, however, because of large mismatch and feed-line losses involved in transferring power to and from electrically short wires.)
\par
\begin{figure}[h]
\mbox{}\\[-7.mm]
%\begin{center}
%\includegraphics[width =4.0in]{Q-AWPL-Fig0.eps}
% Syntax:  \centerbmp{<width>}{<height>}{<path+filename>}
%     Requires "\input setbmp" at the beginning of your file.
% Optional:  <path> (use / instead of \), specifies path of TeX file if not supplied.
% Example:  \centerbmp{3cm}{4cm}{c:/mysubdir/mypic.bmp}
%\centerbmp{6.0in}{4.0in}{GA-supergain.bmp}
%\seteps{-.12in}{3.75in}{2.5in}{Error_in_two_element3.eps}
\includegraphics[width=3.75in,height=2.5in]{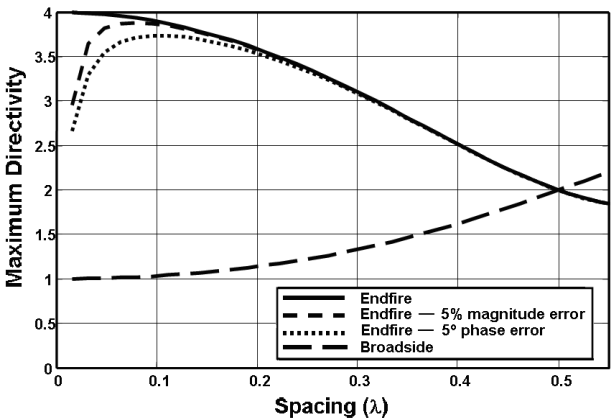}
\mbox{}\\[-15.mm]
%\end{center}
\caption{\label{fig1}Change in maximum directivity versus separation distance of a two-element array of isotropic radiators caused by magnitude and phase errors in the excitation of the first element.}
\end{figure}
\begin{figure}[h]
%\mbox{}\\[-9.mm]
%\begin{center}
%\includegraphics[width =4.0in]{Q-AWPL-Fig0.eps}
% Syntax:  \centerbmp{<width>}{<height>}{<path+filename>}
%     Requires "\input setbmp" at the beginning of your file.
% Optional:  <path> (use / instead of \), specifies path of TeX file if not supplied.
% Example:  \centerbmp{3cm}{4cm}{c:/mysubdir/mypic.bmp}
%\centerbmp{6.0in}{4.0in}{GA-supergain.bmp}
%\seteps{-.04in}{3.65in}{2.45in}{Error_in_three_element3.eps}
\includegraphics[width=3.65in,height=2.45in]{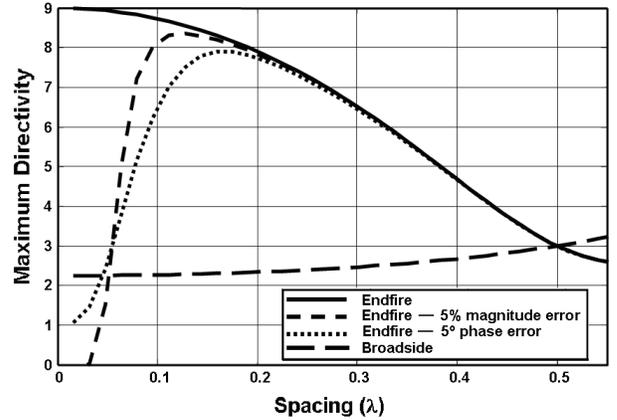}
\mbox{}\\[-15.5mm]
%\end{center}
\caption{\label{fig2}Change in maximum directivity versus separation distance of a three-element array of equally spaced isotropic radiators caused by magnitude and phase errors in the excitation of the first element.\vspace{-4mm}}
\end{figure}

As part of our own error analyses, we calculated the maximum endfire directivities versus element spacing with either a 5\% magnitude error or a $5^\circ$ phase error in the excitation coefficient of the first element of a two-element and three-element endfire array \cite{Altshuler-2005}.  The results shown in Figs. \ref{fig1} and \ref{fig2} for electrically small isotropic radiators reveal that these magnitude and phase errors do not decrease the maximum endfire directivity ($N^2$) by more than about 10\% for two- and three-element arrays if the spacing of the array elements is larger than about $0.05 \lambda$ and $0.15 \lambda$, respectively. Moreover, at separation distances of about $0.05 \lambda$ and $0.15 \lambda$, these two- and three-element arrays of isotropic radiators have directivities close to their maximum possible values of $N^2 =4$ and 9, respectively.  The maximum broadside directivities of these arrays are also shown in Figs.  \ref{fig1} and \ref{fig2} for the sake of comparison with the endfire directivities. {\em Electrically small} broadside arrays of $N$ equally spaced isotropic radiators cannot produce a gain greater than $N$.
%\mbox{}\\[-5mm]
%
%\mbox{}\\[-5mm]
\par
%\mbox{}\\[-5mm]
To recapitulate, although tolerance constraints prevent the practical realization of significant supergain for endfire arrays with more than a few elements, calculations show that the maximum possible endfire gains of arrays with two, three, and possibly more elements can be approached without encountering prohibitive tolerance constraints.  Also, for endfire supergain arrays where beam steering is not required, the strong mutual coupling between the closely spaced elements does not have to be reduced in order to properly drive the elements, as would be the case for broadside steered-beam superdirective arrays \cite{Buell-Sarabandi}.
\subsection{Input-Impedance Mismatches}
An electrically small time-harmonic ($e^{j\w t}$, $\w =2\pi f >0$) antenna operating at a frequency $f$ well below its first resonant or antiresonant\footnote{By definition, an antenna operates at a resonant or antiresonant frequency $f$ if its input reactance $X(f)$ is zero and
 $dX(f)/df >0$ or $dX(f)/df <0$, respectively, \cite{Y&B}.} frequency is generally either a capacitive electric dipole with a reactance that behaves as $1/f$ and a radiation resistance that behaves as $f^2$, or an inductive magnetic dipole with reactance that behaves as $f$ and a radiation resistance that behaves as $f^4$ \cite{B&Y}. This extremely low radiation resistance of a magnetic-dipole antenna operating well below its first resonance makes it unsuitable for use as an element in an efficient antenna array, and thus we are left with only electrically small electric-dipole elements in the class of antennas that can be used in supergain arrays well below their first resonance.
\par
However, the high capacitive reactance of below-resonance electric-dipole elements generally requires cancelation by tuning inductive reactances in order to feed the antenna array a reasonable amount of power.  For example, an electric dipole operating at one-third its resonant frequency typically has a negative input reactance of more than 1200 ohms and a radiation resistance of about 6 ohms \cite[fig. 3]{Y&B}.  Depending on the frequency, a 1200 ohm tuning inductor may add an appreciable ohmic loss to the electric-dipole element and significantly increase its size without increasing its radiation resistance.  An alternative to tuning a highly reactive,  below-resonance, electrically small antenna element is to use a self-resonant antenna element having the same electrical size.  This alternative yields an antenna element with negligible input reactance while keeping the ohmic losses to a minimum.  In addition, as we explain in the next subsection, electrically small resonant antennas can be designed with high radiation resistances and efficiencies, at least at and below GHz frequencies.
\subsection{Radiation Efficiency}
The radiation resistance of an electrically short, straight-wire, electric dipole antenna of length $2a$ has a radiation resistance given by $20 (ka)^2$ ohms \cite[p. 176, eq. (11)]{Kraus-book}. Simulations with the Numerical Electromagnetics Code (NEC) \cite{NEC} indicate that a well-designed electrically small, open-ended (as opposed to closed-loop or folded), bent-wire resonant antenna can have a radiation resistance of 2 to 3 times this value \cite{Altshuler-GA}.  Thus, as a rough approximation, we can assume a radiation resistance of
\be{rad}
R_{\rm rad} \approx 50 (ka)^2 \mbox{ ohms}
\ee
for a well-designed, electrically small, open-ended, bent-wire resonant antenna.  For example, if $a/\lambda = 1/20$ ($ka =0.314$), then $R_{\rm rad} \approx 5$ ohms.  
%A typical open-ended (as opposed to closed-loop or folded) wire antenna with a $ka \approx .5$  resonating at a frequency of about $f =400$ MHz has a radiation resistance in free space of $R_{\rm rad}\approx 10$ ohms \cite{Best-January2005}. (This free-space value of 10 ohms is twice the approximate value of the 5 ohms given in \cite{Best-January2005} for resonant antennas over a perfectly electrically conducting ground plane.)
\par  
A well-designed, electrically small,  open-ended, bent-wire antenna resonant in free-space would have an overall length of $l \approx \lambda/2$.  If the wire material is copper with a conductivity of $\sigma = 5.75 \times 10^7$ mhos/m, and the cross section of the wire is circular with a diameter of $d = 1.6$ mm, the loss resistance $R_{\rm loss}$ of an antenna resonating at a frequency of about $f =400$ MHz is given approximately by the formula \cite[eq. (4)]{B&Y}
\be{1}
R_{\rm loss} \approx \frac{l}{2 d} \sqrt{\frac{f \mu_0^{}}{\pi\sigma}}\;
\stackrel{l\approx \lambda/2}{\approx} \frac{Z_0}{4 d} \sqrt{\frac{1}{\pi \mu_0^{}\sigma f}}
\ee
where $\mu_0^{}$ and $Z_0$ are the permeability and impedance of free space, respectively.\footnote{Note that for an electrically short, straight-wire electric dipole, $l=2a$ and 
\be{1a}
R_{\rm loss} 
\stackrel{l=2a}{\approx} \frac{ka Z_0}{2\pi d} \sqrt{\frac{1}{\pi \mu_0^{}\sigma f}}
\ee
so that at a fixed value of $ka$ the loss resistance of both an electrically short, straight-wire electric dipole and a well-designed, electrically small, open-ended, bent-wire resonant antenna varies as $1/ (d\sqrt{\sigma f})$.}  For the values of the parameters given above, we have $R_{\rm loss} \approx 0.2$ ohms.  The radiation efficiency $\eta$ of this well-designed, electrically small, open-ended, bent-copper-wire resonant antenna is then given by
\be{2}
\eta =  \frac{R_{\rm rad}}{R_{\rm rad} + R_{\rm loss}}  \approx \frac{5.0}{5.2} \approx96 \%\,.
\ee
\par
\begin{figure}[h]
\mbox{}\\[-15mm]
%\begin{center}
%\includegraphics[width =4.0in]{Q-AWPL-Fig0.eps}
% Syntax:  \centerbmp{<width>}{<height>}{<path+filename>}
%     Requires "\input setbmp" at the beginning of your file.
% Optional:  <path> (use / instead of \), specifies path of TeX file if not supplied.
% Example:  \centerbmp{3cm}{4cm}{c:/mysubdir/mypic.bmp}
%\centerbmp{6.0in}{4.0in}{GA-supergain.bmp}
%\seteps{-.33in}{4.20in}{2.92in}{Powerfig2.eps}
\includegraphics[width=4.20in,height=2.92in]{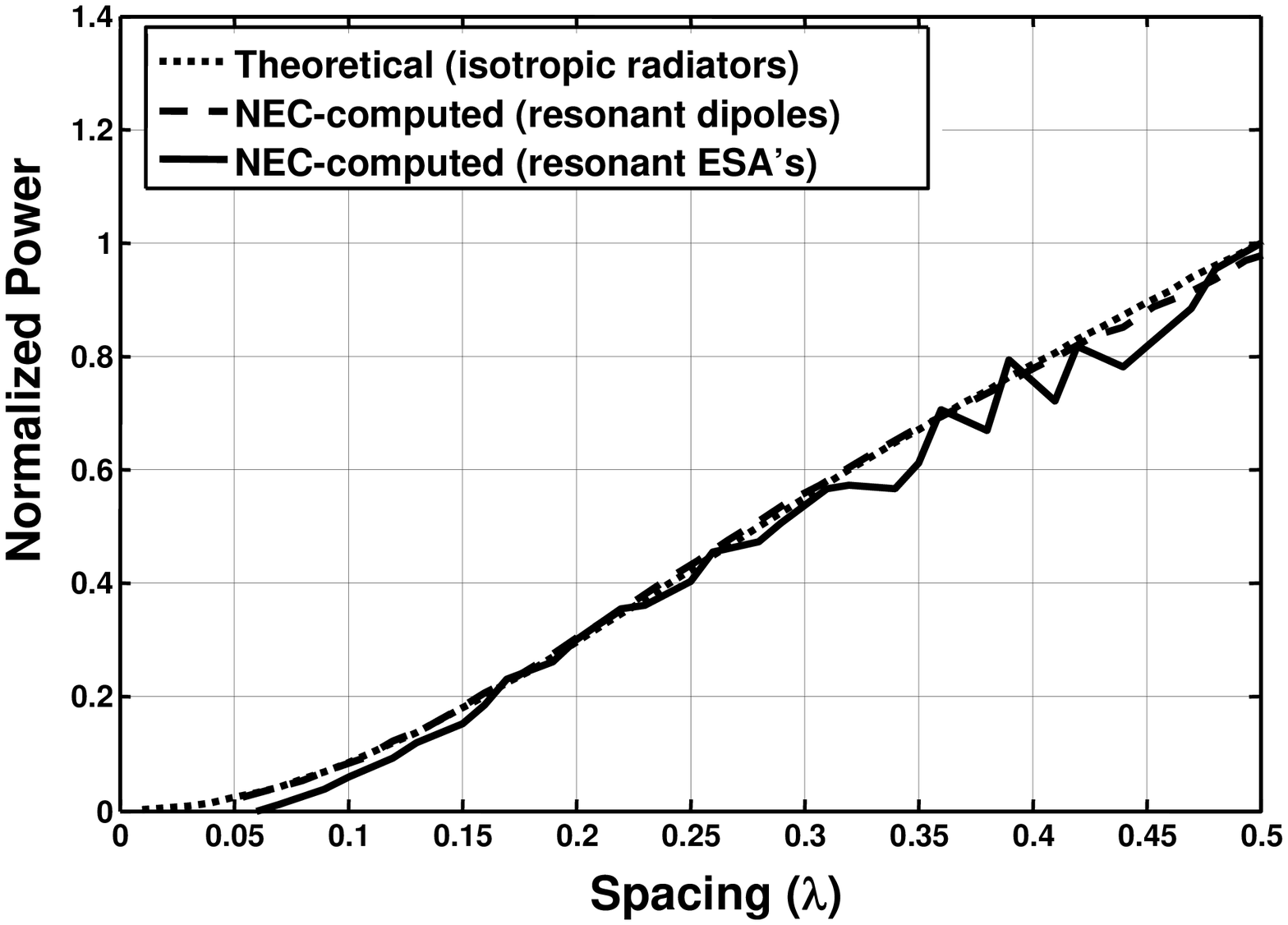}
\mbox{}\\[-19.5mm]
%\end{center}
\caption{\label{fig3}Normalized power for two-element superdirective arrays of isotropic radiators, resonant electric dipoles, and resonant electrically small antennas.\vspace{-1mm}}
\end{figure}

This high radiation efficiency of $96\%$ in (\ref{2}) for a single electrically small resonant antenna will be reduced, however, as two of these antennas are brought close together and are properly fed to produce a two-element supergain array.  The radiation efficiency of such a two-element array can be estimated with the help of the power curve shown
 in Fig. \ref{fig3}.   This curve shows the total power radiated by two isotropic (acoustic) radiators, two resonant straight-wire electric dipoles, and two resonant electrically small  antennas (ESA's), each pair fed with unity magnitude current having the proper phase difference to produce the maximum endfire directivity at a given separation distance.  The total power is normalized to unity at a spacing of $.5\lambda$. (For two-element arrays, the maximum directivity at any spacing is attained with equal current magnitude fed to each element.)  The main reason for the decrease in power radiated is that, as the elements get closer, the phase difference between the equal magnitude currents approaches $180^\circ$.  Thus, the fields produced by these currents tend to cancel and the array element radiation resistance decreases in proportion to the normalized power shown in Fig. \ref{fig3}.  At a separation distance of $0.15\lambda$, the normalized radiated power in Fig. \ref{fig3}, and thus the radiation resistance of the two-element array, is about 0.2 times the radiation resistance of the individual elements in free space.  Consequently, the 5 ohm radiation resistance of the aforementioned 400 MHz ($a/\lambda =1/20, ka = 0.314$) resonant antenna would be reduced to about 1 ohm for two of these antennas separated by $0.15\lambda$ and the radiation efficiency of this two-element array would be reduced to about $\eta =83 \%$, which represents a reduction in the supergain of about 0.8 dB.
%\mbox{}\\[-5mm]
An ohmic-loss reduction of about 0.8 dB or less in the 6 to 7 dB maximum endfire gain of an electrically small two-element array does not compromise the supergain to a great extent.  And, in fact, the first two-element supergain array that we measured to confirm that a supergain close to the maximum predicted value of 6 to 7 dB could be achieved experimentally was constructed from two electrically small ($a/\lambda \approx 1/18, ka \approx 0.35$), open-ended, bent-copper-wire antennas resonant at about 400 MHz with a free-space radiation resistance of about 6 ohms (reducing to about 1.2 ohms at a separation of $0.15 \lambda$); see Section \ref{sec III-A}.
\par
Although low-loss miniature matching circuits have been designed to feed electrically small antennas with radiation resistances less than 1 ohm \cite{Yoshida}, it nonetheless introduces additional difficulties  to efficiently feed an antenna with input resistances of a few ohms or less.  Fortunately, for electrically small, open-ended, bent-wire resonant antennas, the radiation resistance can be greatly increased simply by adding a small tuning loop (or post) across the feed point in parallel with the original antenna \cite{Best-January2005}--\cite{Lim-Ling-2004}. A small tuning loop provides the main conduction path for the resonant current and thus lowers the feed-point current for a given applied voltage, thereby increasing the input resistance.  It does not, however, significantly change the radiation efficiency or bandwidth because the stored energy and power radiated is still determined predominantly by the resonant current on the original bent-wire antenna.  Thus,  tuning loops alleviate the problem of matching to a very low radiation resistance but they do not increase the radiation efficiency of electrically small, open-ended, bent-wire resonant antennas.\footnote{Electrically small, low-loss, wire-loop antennas operating at their first antiresonant frequency have  radiation resistances too high (usually many thousands of ohms) to feed without sophisticated circuitry that would increase the size and lower the efficiency of such antennas. One can excite a wire-loop antenna at a frequency slightly above or below the antiresonant frequency, then retune the antenna to zero reactance with an inductor or capacitor, respectively, to obtain a much lower input resistance (50 ohms, for example). (A similar technique has been used to match the impedance of slot antennas \cite{Sarabandi}.)  Unfortunately, this matching technique  does not also increase the radiation efficiency and it decreases the bandwidth of the wire-loop antenna.}
\par
An approach that increases both the radiation resistance and efficiency of resonant antennas, including electrically small resonant antennas, is to use multiple folded arms \cite{Dobbins&Rogers}--\cite{Lim-Ling-2006}.  The half-wavelength, straight-wire, folded dipole is the classic example of such a resonant antenna (although it is not electrically small) \cite[sec. 9.5]{Balanis}, but any number of bent-wire folded resonant antenna designs display the same attractive features of a higher radiation resistance combined with a higher radiation efficiency and often a greater bandwidth (lower $Q$).  In its essence, an electrically small, bent-wire, folded resonant antenna with $M$ arms (including the feed arm) is a loop antenna with $M-1$ bent wires connecting the top and bottom of the bent-wire arm that is fed. With a symmetric design all of the $M$ arms carry approximately the same resonant current as the feed arm and thus the total power radiated by the antenna scales approximately as $M^2$.  The antenna's ohmic loss resistance, however, scales approximately only as the number of arms $M$ \cite{Lim-Ling-2006} and thus the efficiency of the antenna increases with $M$ as 
\be{M}
\eta = \frac{M^2}{M^2 +\alpha M} =\frac{1}{1+\alpha/M}
\ee
which approaches unity as $M$ gets large (until the number of arms and bends start to interfere with one another).  The constant $\alpha$, which is proportional to the resistivity of the wire material, can be expressed in terms of the efficiency $\eta_1$ of the original one-arm (M=1) bent-wire antenna by the formula $\alpha =1/\eta_1 -1$.
\par
We employed a combination of bends, folds, and tuning posts in NEC to design efficient, electrically small, bent-wire, resonant antennas with appreciable radiation resistances and reasonably low values of $Q$.  These resonant antennas can then be used as the elements in electrically small, separately fed and singly fed (parasitic), two-element, supergain endfire arrays \cite{O&Y}; see the next subsection and Section \ref{sec III-B}.
\subsection{Parasitic Elements}
\begin{figure}[h]
%\mbox{}\\[-4.5mm]
%\begin{center}
%\includegraphics[width =4.0in]{Q-AWPL-Fig0.eps}
% Syntax:  \centerbmp{<width>}{<height>}{<path+filename>}
%     Requires "\input setbmp" at the beginning of your file.
% Optional:  <path> (use / instead of \), specifies path of TeX file if not supplied.
% Example:  \centerbmp{3cm}{4cm}{c:/mysubdir/mypic.bmp}
%\centerbmp{6.0in}{4.0in}{GA-supergain.bmp}
%\seteps{-.05in}{3.63in}{2.36in}{Parasiticmonopole.eps}
\includegraphics[width=3.63in,height=2.36in]{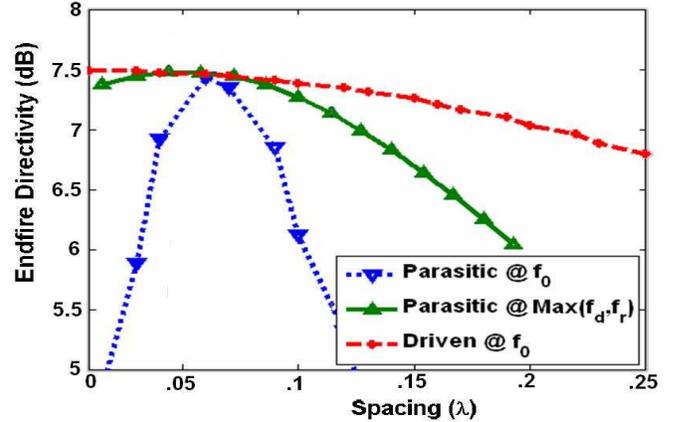}
\mbox{}\\[-13.mm]
%\end{center}
\caption{\label{fig4}Endfire directivity versus separation distance of two nominally half-wavelength, lossless, straight-wire dipoles for three cases: both elements optimally driven to obtain maximum directivities at the individual-element resonant frequency $f_0$; one element shorted and the other driven at the resonant frequency $f_0$; and one element shorted and the other driven at shifted frequencies $f_d$ and $f_r$ that produce maximum directivities in the endfire directions for which the parasitic element is a director or a reflector, respectively.\vspace{-5mm}}
\end{figure}

The maximum endfire directivity versus separation distance of two parallel, separately driven, nominally half-wavelength, 1.6 mm diameter, lossless, straight-wire dipoles is shown by the dashed curve (labeled ``driven") in Fig. \ref{fig4}, where $f_0 = 437$  MHz is the resonant frequency of each of the dipoles in free space. In free space each resonant dipole has an input (radiation) resistance of 72 ohms and a $Q$ of 5.6.  Each of the dipoles is fed with the same current magnitude and with the phase difference determined with NEC that produces the maximum directivity at each separation distance.  As the separation distance approaches zero, the maximum directivity approaches 7.5 dB, a value that is 1.5 dB higher than the maximum directivity of 6 dB ($N^2= 4$) for two isotropic radiators approaching zero separation distance.
\mbox{}\\[-5mm]
\par
If the same two half-wavelength elements are used to form a parasitic (Yagi) antenna with one element fed at the individual resonant frequency of $f_0 = 437$ MHz, and the parasitic element {\em shorted}, the directivity versus separation distance is shown by the dotted curve in Fig. 4.  And if the frequency of the one fed element is shifted slightly (typically not more than a few MHz) to a value $f_d$ or $f_r$ to maximize the directivity at each separation distance, depending on whether the maximum occurs with the shorted parasitic dipole acting as a director (subscript ``{\em d}") or a reflector (subscript ``{\em r}"), the maximum directivity versus separation distance is shown by the solid curve in Fig. \ref{fig4}. (The direction of maximum directivity switches from the parasitic dipole acting as a reflector to the parasitic dipole acting as a director at a separation distance of about $0.12 \lambda$.) The most important feature of the two parasitic curves in Fig. \ref{fig4} is that they reach a maximum directivity (which always occurs when the array is a driver-director Yagi) greater than 7.4 dB, that is, less than $0.1$ dB below the highest possible maximum of 7.5 dB for the separately driven elements.  (This maximum value of about 7.4 dB compares quite well with the theoretical estimate of about 7.3 dB for the maximum directivity of a driver-director Yagi obtained by Walkinshaw \cite[fig. 2(a)]{Walkinshaw}.) Moreover, if the loss of the copper wire is taken into account in the NEC code, the maximum gain that is reached for the separately fed and parasitic two-element arrays is about 7.25 dB; that is, the difference between the maximum possible directivity and gain of the two dipoles is about 0.25 dB. At $0.1\lambda$ separation distance, the NEC-computed gain of the lossy two-element Yagi is 7.17 dB, its efficiency is 97.6\%, its input impedance is $13.4 - 29.6i$ ohms, and its $Q$ is 53.8 after  tuning the negative 29.6 ohm reactance to zero with a small series inductor.  This value of $Q$ corresponds to a 3.7\% matched voltage-standing-wave-ratio (VSWR) half-power fractional bandwidth \cite{Y&B}.
\par
There are two main reasons why two closely spaced, nominally half-wavelength, straight-wire dipoles form a parasitic array (Yagi) that achieves nearly the same maximum possible gain as two separately (and optimally) fed closely spaced half-wavelength straight-wire dipoles.
\par
First, the {\em shorted} parasitic element forms a resonant dipole scatterer, so that for closely spaced and thus strongly coupled elements, the magnitude of the current on the parasitic element can be as large as that on the driven element.  Second, the phase difference between the resonant current on the driven and parasitic elements is close to $180^\circ$. In other words, the closely spaced, two-element, resonant Yagi is operating predominantly in the odd mode of two coupled resonators.
\par
Since the directivity of two closely spaced antennas is maximized if the magnitudes of the currents on each element are equal and the phase difference between the currents is close to $180^\circ$ (see \cite[fig. 8]{Altshuler-2005}), it follows that on either side of the resonant frequency at which nearly equal magnitude currents are nearly $180^\circ$ out of phase, approximately the maximum possible directivity is attained.
%Our NEC4 simulations of two coupled electrically small resonant antennas, one fed and the other shorted, confirmed this conjecture.  In fact, as might be expected, two relative maxima in the directivity are found, one at a frequency a little higher and another at little lower than the new resonant frequency where the elements are nearly 180o out of phase.
 One relative maximum corresponds to the endfire direction of a driver-director parasitic (Yagi) array and the other to the endfire direction of a driver-reflector parasitic (Yagi) array.
\par
In view of these foregoing two reasons why two-element parasitic arrays of closely spaced, nominally half-wavelength, straight-wire, resonant dipoles can attain such a high directivity, it is apparent why shortening the wires to make them electrically small would eliminate the possibility of high directivity.  The magnitude of the current on an electrically short, straight-wire, parasitic element, not being resonant, would not be nearly as large as on the driven element (without heavy impedance loading on the driven element \cite{Nyquist}), and the coupling between these two below-resonance elements would not necessarily produce the approximate $180^\circ$ phase change between the currents on the elements that is required to produce supergain.
\par
Upon examination of these two foregoing reasons for high-gain, half-wavelength, two-element, parasitic (Yagi) arrays, it also becomes apparent that they can apply to numerous resonant antenna elements regardless of their electrical sizes. {\em In other words, there appears to be no reason why two electrically small resonant antennas could not be used as elements in an electrically small two-element supergain array in which one resonant antenna was driven and the other resonant antenna was shorted to form a resonant scatterer.}  And, indeed, NEC-computed simulations with numerous two-element parasitic arrays of electrically small resonant antennas verified this conjecture.
\par
\begin{figure}[h]
\mbox{}\\[-1.5mm]
%\begin{center}
%\includegraphics[width =4.0in]{Q-AWPL-Fig0.eps}
% Syntax:  \centerbmp{<width>}{<height>}{<path+filename>}
%     Requires "\input setbmp" at the beginning of your file.
% Optional:  <path> (use / instead of \), specifies path of TeX file if not supplied.
% Example:  \centerbmp{3cm}{4cm}{c:/mysubdir/mypic.bmp}
%\centerbmp{6.0in}{4.0in}{GA-supergain.bmp}
%\seteps{-.25in}{3.95in}{2.20in}{ParasiticTest55.eps}
\includegraphics[width=3.95in,height=2.20in]{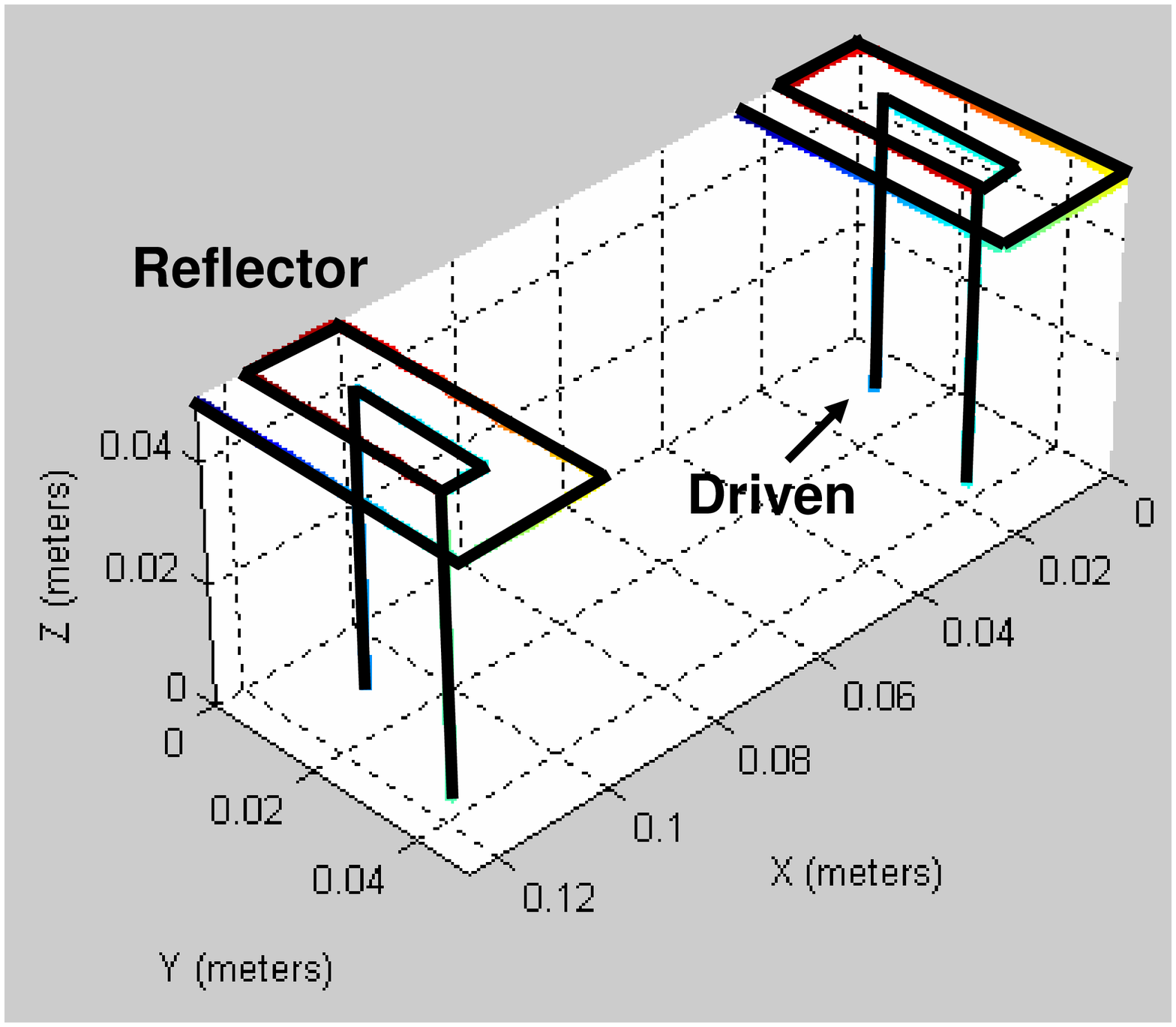}
\mbox{}\\[-11.mm]
%\end{center}
\caption{\label{fig5}Two electrically small ($ka = 0.5$ in free space), top-loaded,  folded, 1.6 mm diameter bent-wire antennas individually resonant at $f_0 =437$ MHz forming a two-element parasitic array.\vspace{-4mm}}
\end{figure}

As an example, two identical electrically small ($ka = 0.5$ in free space), top-loaded,  folded, 1.6 mm diameter bent-wire antennas individually resonant at $f_0 =437$ MHz form the two-element parasitic array (one element driven and the other shorted) shown in Fig. \ref{fig5}. (Fig. \ref{fig5} shows the array over an infinite perfectly electrically conducting (PEC) ground plane.   In free-space, the array would include the image of the elements.) Each of the resonant antennas alone over ground has a radiation resistance of 61 ohms and a $Q$ of 38; in free-space the corresponding resonant antennas have a radiation resistance of 122 ohms and the same $Q =38$.
%\mbox{}\\[-5mm]
%
\par
\begin{figure}[h]
\mbox{}\\[-5.mm]
%\begin{center}
%\includegraphics[width =4.0in]{Q-AWPL-Fig0.eps}
% Syntax:  \centerbmp{<width>}{<height>}{<path+filename>}
%     Requires "\input setbmp" at the beginning of your file.
% Optional:  <path> (use / instead of \), specifies path of TeX file if not supplied.
% Example:  \centerbmp{3cm}{4cm}{c:/mysubdir/mypic.bmp}
%\centerbmp{6.0in}{4.0in}{GA-supergain.bmp}
%\seteps{0.in}{3.54in}{2.38in}{ParasiticTest55a2.eps}
\includegraphics[width=3.54in,height=2.38in]{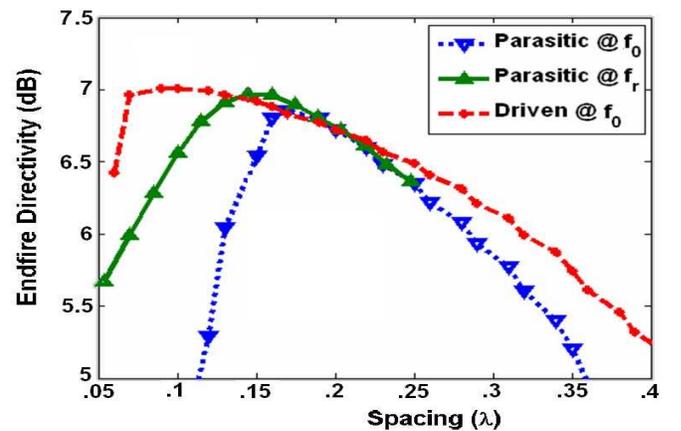}
\mbox{}\\[-14.mm]
%\end{center}
\caption{\label{fig6}Endfire directivity versus separation distance of the electrically small two-element array in Fig. \ref{fig5} for three cases (with the wire  lossless and 3 dB subtracted because of the ground plane): both elements optimally driven to obtain maximum directivities at the individual-element resonant frequency $f_0$; one element shorted and the other driven at the resonant frequency $f_0$; and one element shorted and the other driven at the shifted frequencies $f_r$ that produce the maximum endfire directivities, which for all separation distances occur in the endfire direction with the parasitic element acting as a reflector.\vspace{-.5mm}}
\end{figure}

The NEC-computed endfire directivity versus separation distance of this two-element parasitic array is plotted in Fig. \ref{fig6}, where 3 dB has been subtracted to give the free-space directivity of the elements (with their image) in the absence of the ground plane. As in Fig. \ref{fig4} for the half-wavelength dipoles, three curves are shown in Fig. \ref{fig6}: the maximum directivities for both elements driven with their optimal excitations; the parasitic directivities at the individual resonant frequency $f_0 = 437$; and the parasitic directivities maximized at each separation distance by shifting the frequency to a value $f_r$.  (Unlike the two-element half-wavelength dipole array, the maximum directivity at all separation distances of this two-element electrically small parasitic array occurs in the endfire direction for which the parasitic element is a reflector rather than a director.)
%\mbox{}\\[-5mm]
%
\par
The curves in Fig. \ref{fig6} reveal the remarkable result that at a separation distance of about $0.15 \lambda$, the parasitic array reaches a maximum directivity that is less than 0.1 dB below the maximum possible separately driven directivity of 7.0 dB. With loss in the copper wires taken into account, the NEC code predicts that  the maximum gain drops slightly to 6.5 dB.   At about a $0.15 \lambda$ separation, the efficiency of the array is about 90\%, its free-space input impedance is about $50 +70i$ ohms, and its $Q \approx 154$  (half-power matched VSWR impedance fractional bandwidth of about 1.3\%)  after tuning out the 70 ohm reactance with a small capacitor. The array also exhibits a 1.3\% fractional bandwidth with respect to a 1 dB drop in gain. The entire two-element array in free space fits into a sphere of $ka \approx 0.7$.  The NEC computations for many other two-element arrays formed with various electrically small folded bent-wire antenna elements produce similar results. 
\section{Experimental Results} \label{sec III}
Two sets of measurements of supergain are given in this section containing the experimental results: the first set is the measured gain versus separation distance of two separately fed, electrically small, open-ended, bent-wire resonant antennas (see Fig. \ref{fig7}), and the second set is the measured gain versus separation distance of two electrically small, planar, doubly folded, bent-wire, resonant antennas (see Fig. \ref{fig10}) fed as a two-element parasitic array. We used these latter two planar antennas rather than the ones shown in Fig. \ref{fig5} because they were easier to build and they allowed us to measure the gain at very close separation distances. All of the antenna wires are made of copper and have a diameter of 1.6 mm. Relative gain measurements were made on a ground plane with respect to the gain of a resonant, nominally quarter-wavelength monopole over the ground plane.  Absolute gain was then determined by adding the computed 2.15 dB gain (plus 3 dB to account for the ground plane) of the resonant quarter-wavelength monopole to the values of the relative gain.
\subsection{ Measured Gain of Separately Fed, Electrically Small, Two-Element Array}\label{sec III-A}
\begin{figure}[h]
\mbox{}\\[-2.5mm]
%\begin{center}
%\includegraphics[width =4.0in]{Q-AWPL-Fig0.eps}
% Syntax:  \centerbmp{<width>}{<height>}{<path+filename>}
%     Requires "\input setbmp" at the beginning of your file.
% Optional:  <path> (use / instead of \), specifies path of TeX file if not supplied.
% Example:  \centerbmp{3cm}{4cm}{c:/mysubdir/mypic.bmp}
%\centerbmp{6.0in}{4.0in}{GA-supergain.bmp}
%\seteps{0.mm}{3.4in}{1.9in}{ParasiticTest55a2.eps}
\includegraphics[width=3.4in,height=1.9in]{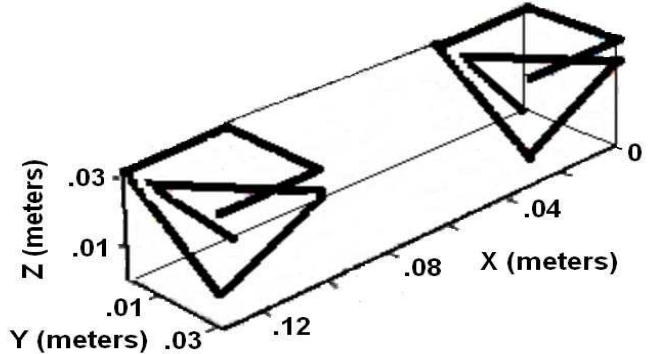}
\mbox{}\\[-6.mm]
%\end{center}
\caption{\label{fig7}Two-element supergain array over an infinite $xy$ PEC ground plane with each element an optimally driven electrically small, seven-segment,  open-ended, bent-copper-wire antenna resonant at about 400 MHz.\vspace{-4mm}}
\end{figure}

Each of the identical antenna elements over the infinite $xy$ PEC ground plane shown in the two-element array of Fig. \ref{fig7} are electrically small, seven-segment,  open-ended, bent-copper-wire antennas resonant at about 400 MHz. Along with its image each antenna element has a free-space value of 
$a/\lambda \approx 1/18, ka \approx 0.35$.  The NEC simulations with loss in the copper wire predict that each of the free-space antennas have a radiation resistance of 5.4 ohms, an efficiency of 94\%, and a $Q$ of 95 (half-power matched VSWR fractional bandwidth of about 2\%). 
%\mbox{}\\[-5mm]
%
\par
\begin{figure}[h]
\mbox{}\\[-4.mm]
%\begin{center}
%\includegraphics[width =4.0in]{Q-AWPL-Fig0.eps}
% Syntax:  \centerbmp{<width>}{<height>}{<path+filename>}
%     Requires "\input setbmp" at the beginning of your file.
% Optional:  <path> (use / instead of \), specifies path of TeX file if not supplied.
% Example:  \centerbmp{3cm}{4cm}{c:/mysubdir/mypic.bmp}
%\centerbmp{6.0in}{4.0in}{GA-supergain.bmp}
%\seteps{1.mm}{3.38in}{2.25in}{GA-supergain-2a.eps}
\includegraphics[width=3.38in,height=2.25in]{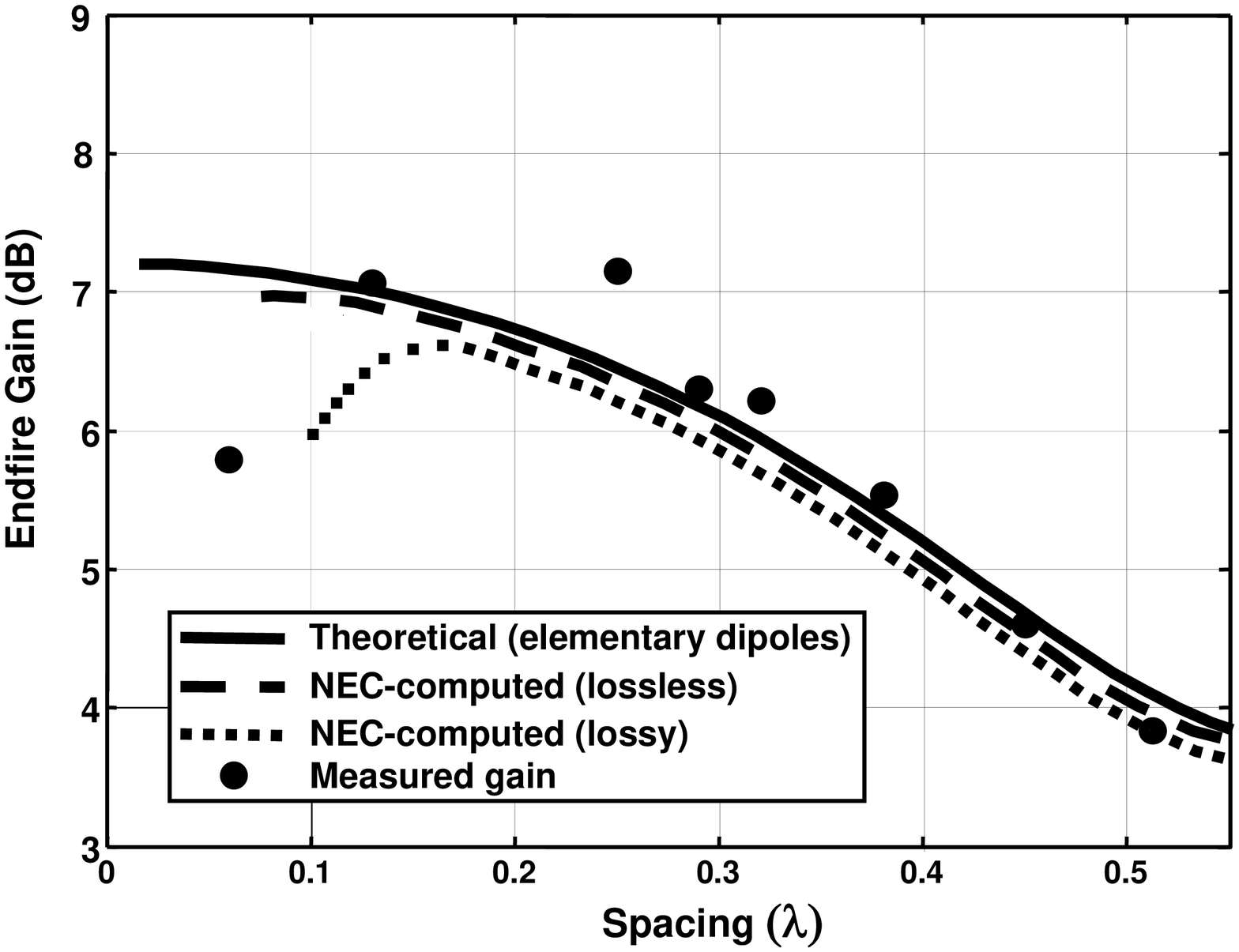}
\mbox{}\\[-7.mm]
%\end{center}
\caption{\label{fig8}NEC-computed and measured maximum endfire gains as a function of separation distance of separately fed two-element array shown in Fig. \ref{fig7} (with 3 dB subtracted because of the ground plane), as well as the maximum theoretical gain of two separately fed elementary dipoles versus separation distance.\vspace{-1mm}}
\end{figure}
\begin{figure}[h]
\mbox{}\\[-1.mm]
%\begin{center}
%\includegraphics[width =4.0in]{Q-AWPL-Fig0.eps}
% Syntax:  \centerbmp{<width>}{<height>}{<path+filename>}
%     Requires "\input setbmp" at the beginning of your file.
% Optional:  <path> (use / instead of \), specifies path of TeX file if not supplied.
% Example:  \centerbmp{3cm}{4cm}{c:/mysubdir/mypic.bmp}
%\centerbmp{6.0in}{4.0in}{GA-supergain.bmp}
%\seteps{1.mm}{3.45in}{1.8in}{Schematic.eps}
\includegraphics[width=3.45in,height=1.8in]{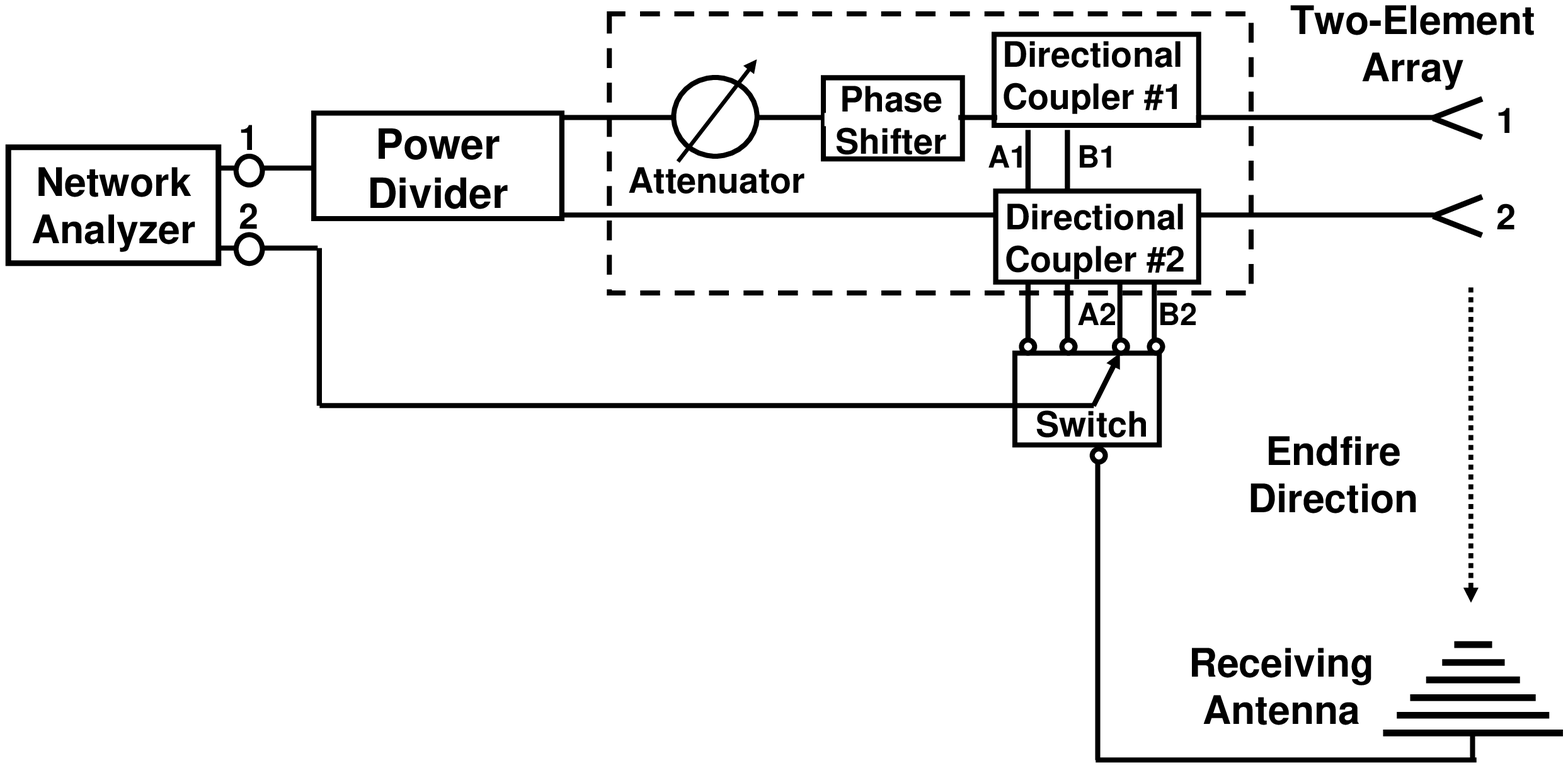}
\mbox{}\\[-7.mm]
%\end{center}
\caption{\label{fig9}Schematic of gain measurement system for two separately fed array elements.\vspace{-5mm}}
\end{figure}

With each of the antenna elements in Fig. \ref{fig7} driven separately at the individual-element resonant frequency and with the optimum currents (equal magnitude and a phase difference given approximately in \cite[fig. 5 (elementary monopole)]{Altshuler-2005}) to produce the maximum endfire directivity, the NEC computations of gain as a function of separation distance are shown in Fig. \ref{fig8} with and without loss in the copper wire.  Also, shown in Fig. \ref{fig8} are the measured values of maximum gain versus separation distance obtained over a finite ground plane with the measurement system depicted schematically in Fig. \ref{fig9}; see also \cite{Altshuler-2005}. Although all the computations and measurements of this two-element array were made over a PEC ground plane, the values of gain in Fig. \ref{fig8} have been reduced by 3 dB to those of the corresponding free-space two element array (comprised of the antennas in Fig. \ref{fig7} and their images in the ground plane).
\par
The curve in Fig. \ref{fig8} of the NEC-computed data for the lossy two-element array of separately fed elements shows that a gain of about 6.7 dB is attained at a separation distance of about $0.15\lambda$, where the entire free-space array fits into a sphere with electrical size of about $ka = 0.7$. This high value of gain, which is just 0.3 dB less than the maximum possible lossless NEC-computed supergain of about 7 dB for these electrically small, open-ended, bent-wire antenna elements, is confirmed by the values of the measured gain shown in Fig. \ref{fig8}.  The solid curve in Fig. \ref{fig8} demonstrates that the theoretically determined values of maximum endfire directivity for two optimally driven elementary dipoles are very close to the gain values computed for the two-element array of optimally driven lossless, electrically small, bent-wire elements.
%\mbox{}\\[-5mm]
%  
%\mbox{}\\[-5mm]
%
\par
Accurate measured values of gain (shown in Fig. \ref{fig8}) were difficult to obtain especially at the smaller separation distances because the initial low value of the input resistance  of each of these bent-wire  elements ($5.7/2 = 2.9$ ohms over the ground plane) decreased  with decreasing separation distance.  This produced a reflected power that was nearly as large as the incident power and thus the accepted power could not be accurately measured with the network analyzer. Repeated measurements indicated that it is unlikely that the values of the measured gain given in Fig. \ref{fig8} have error bars less than about $\pm 1$ dB for separation distances of less than about $0.25\lambda$. Nonetheless, these early measurements strongly indicated that  values of supergain of between 6 and 7 dB could indeed be obtained  with separately (and optimally) driven, electrically small, two-element, bent-copper-wire arrays.  Of course, we could have made additional, more accurate measurements with separately driven electrically small elements that have much higher input radiation resistances (such as those shown in Fig. {\ref{fig5}), but our discovery that parasitic (single feed), electrically small, two-element arrays exhibited practically the same supergain as with separately driven array elements led us to abandon the tedious procedure required for the gain measurement of separately driven two-element arrays.
\subsection{Measured Gain of Electrically Small Parasitic Two-Element Array}
\label{sec III-B}
A single element (a planar doubly folded bent-copper-wire antenna) of the two-element parasitic array that we measured is shown in Fig. \ref{fig10}. 
\begin{figure}[h]
\mbox{}\\[-4.mm]
%\begin{center}
%\includegraphics[width =4.0in]{Q-AWPL-Fig0.eps}
% Syntax:  \centerbmp{<width>}{<height>}{<path+filename>}
%     Requires "\input setbmp" at the beginning of your file.
% Optional:  <path> (use / instead of \), specifies path of TeX file if not supplied.
% Example:  \centerbmp{3cm}{4cm}{c:/mysubdir/mypic.bmp}
%\centerbmp{6.0in}{4.0in}{GA-supergain.bmp}
%\seteps{-2mm}{3.55in}{2.3in}{parasiticTest131.eps}
\includegraphics[width=3.55in,height=2.3in]{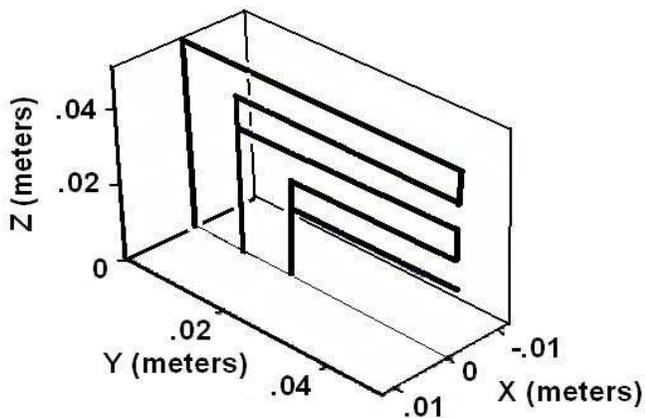}
\mbox{}\\[-11.mm]
%\end{center}
\caption{\label{fig10}Electrically small, planar, doubly folded, bent-copper-wire  antenna resonant at about 876 MHz used in measured two-element parasitic array.}
\end{figure}
%\mbox{}\\[-5mm]
%
The two elements were oriented parallel to each other and separated along the normals to their planes. The NEC-computations and measurements were done over a ground plane with the driven element fed at $(x,y,z) =(0,0,0)$ and the parasitic element shorted at its feed point.  Each of the antennas fed alone has a resonant frequency of about 876 MHz and, along with its image in free space, each has a circumscribing sphere of electrical size $ka \approx 1$.  Each antenna element has a $Q$ of about 4.3, a radiation resistance in free space of about 284 ohms, and a radiation efficiency greater than 99.5\%.
\par
For small fractional wavelength separations, the two-element array of these planar antennas also has a $ka \approx 1$.  Although this borders on being electrically small, the high radiation resistance, high efficiency, and low $Q$ of these planar array elements allowed for more accurate measurements.  Still, the edge effects of the finite ground plane (4 feet by 4 feet), on which the measurements were made, introduced error bars estimated at $\pm 0.5$ dB.  
\par
The NEC-computed and measured endfire gains versus separation distance of this two-element parasitic array are plotted in Fig. \ref{fig11}. At each separation distance, the frequency was shifted to obtain the maximum endfire gain, which was always in the direction with the parasitic element acting as a reflector rather than as a director.  Figure \ref{fig11} shows that the highest maximum values of the NEC-computed and measured gains of the lossy parasitic array in free space occur between the separation distances of $0.05\lambda$ and $0.12\lambda$.  In particular, the maximum computed and measured values of endfire gain are both equal to about 7 dB (with 3 dB subtracted from their ground-plane values) at a spacing of $0.1\lambda$, where the free-space electrical size of the two-element array (with its image) is $ka \approx 1$. This gain value of 7 dB is only about 0.3 dB lower than the maximum attainable value of endfire gain (7.3 dB) as computed with NEC for a two-element array of these planar antenna elements when they are lossless; see solid curve in Fig. \ref{fig11}.  At a separation distance of $0.1\lambda$, the maximum endfire gain is obtained at a frequency of about 874 MHz, the efficiency of the array is about 98.5\%, its free-space input impedance is about $61 + 118i$ ohms, and its value of $Q$ is about 41 (half-power matched VSWR impedance fractional bandwidth of about 4.8\%) after tuning out the 118 ohm reactance with a small capacitor.  The array exhibits about an 8\% fraction bandwidth with respect to a 1 dB drop in gain.
\begin{figure}[h]
\mbox{}\\[-5.mm]
%\begin{center}
%\includegraphics[width =4.0in]{Q-AWPL-Fig0.eps}
% Syntax:  \centerbmp{<width>}{<height>}{<path+filename>}
%     Requires "\input setbmp" at the beginning of your file.
% Optional:  <path> (use / instead of \), specifies path of TeX file if not supplied.
% Example:  \centerbmp{3cm}{4cm}{c:/mysubdir/mypic.bmp}
%\centerbmp{6.0in}{4.0in}{GA-supergain.bmp}
%\seteps{0.mm}{3.37in}{2.35in}{parasiticTest131a.eps}
\includegraphics[width=3.37in,height=2.35in]{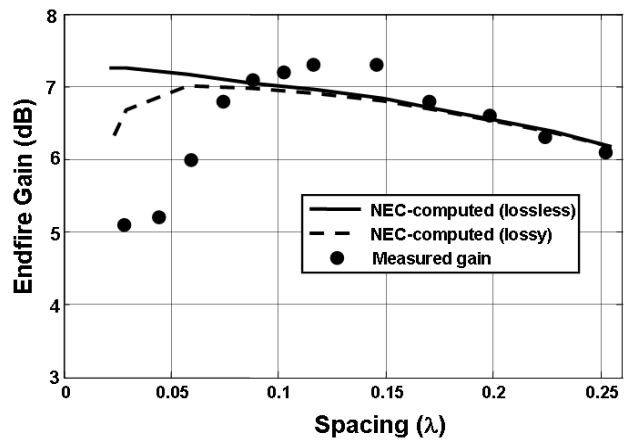}
\mbox{}\\[-6.mm]
%\end{center}
\caption{\label{fig11}NEC-computed and measured maximum endfire gains as a function of separation distance of a parasitic two-element array formed with the antenna element shown in Fig. \ref{fig10} (with 3 dB subtracted because of the ground plane).}
\end{figure}
\mbox{}\\[-0mm]
Clearly, this 7dB-gain array constructed simply from a driver-reflector pair of planar bent-copper-wire resonant antennas demonstrates the feasibility and practicality of producing many other similarly efficient, well-matched, electrically small, two-element, parasitic supergain endfire arrays.
\section{Concluding Remarks} \label{sec IV}
By using resonant antennas as the elements in a two-element array, we have shown from theory, numerical simulation, and experimental measurements that the difficulties of narrow tolerances, large mismatches, low radiation efficiencies, and reduced reflector-element or director-element scattering can be overcome to enable the practical design and construction of {\em electrically small} ($ka < 1$) supergain two-element endfire arrays with gains as high as 7 dB. We showed that this enhanced value of gain, which is just a few tenths of a dB less than the maximum theoretically possible gain of these two-element arrays, could be obtained with one resonant element driven and the other shorted to form a parasitic two-element array as well as with  separately (and optimally) driven resonant elements.  Although rapidly increasing narrow tolerances prevent the practical realization of the maximum theoretically possible endfire gain of electrically small arrays with many elements, the theory and preliminary numerical simulations indicate that near maximum supergains are also achievable in practice for electrically small arrays with three and four resonant elements, and possibly, though less likely, with more than four resonant elements.
\par
The half-power matched VSWR impedance fractional bandwidth of the electrically small supergain two-element parasitic arrays was found from the theory, computations, and measurements to be no more than a few percent. For electrically small arrays with  more than two elements and greater supergains, the bandwidth would be appreciably less.  Thus, the future development of electrically small supergain arrays would naturally entail research into increasing their bandwidth, possibly through the use of electrically small antenna elements with multi-resonances and the incorporation of nonlinear matching networks. Also, future work could involve the use of dielectric and permeable materials to both further reduce the size and confine the high reactive fields of these electrically small antennas, thereby decreasing their susceptibility to interference from nearby structures.
\par
Before concluding, it may be helpful to relate the 7 dB gain of the electrically small two-element supergain arrays to the maximum directivity $D_{\rm max}$ attainable from an antenna that radiates vector spherical multipoles of maximum order $N_{\rm max}$.  For electric multipoles alone or magnetic multipoles alone, it can be proven that \cite{Harrington-book}, \cite{Karlsson}
\be{multi1}
D_{\rm max}^{\rm (e\, or\, m)} =\frac{N_{\rm max}(N_{\rm max} +2)}{2}
\ee  
and for electric and magnetic multipole antennas
\be{multi2}
D_{\rm max}^{\rm (e\, and\, m)} = N_{\rm max}(N_{\rm max} +2)\,.
\ee  
For dipole antennas, $N_{\rm max}=1$,  $D_{\rm max}^{\rm (e\, or\, m)} = 1.5$ (1.76 dB), and $D_{\rm max}^{\rm (e\, and\, m)} = 3.0$ (4.77 dB). For dipole-quadrupole antennas, $N_{\rm max}=2$,  $D_{\rm max}^{\rm (e\, or\, m)} = 4.0$ (6.02 dB), and $D_{\rm max}^{\rm (e\, and\, m)} = 8.0$ (9.03 dB).  Consequently, the far fields of the 7dB-gain electrically small two-element arrays comprise a combination of significant electric and magnetic dipoles and quadrupoles.
%
%\newpage

\end{document}